\documentclass[12pt,a4paper]{article}

\usepackage{amsmath,amssymb,amsfonts,mathtools,bm}
\usepackage{graphicx}
\usepackage{caption}
\usepackage{subcaption}
\usepackage{multirow}
\usepackage{authblk}
\usepackage{setspace}

\title{Musical Attention Transformer: Music Generation Using a Music-Specific Attention Model}
\author[1]{Shinnosuke Takasuka}
\author[1]{Hideo Mukai}
\affil[1]{Department of Computer Science, School of Science and Technology, Meiji University,
1-1-1 Higashimita, Tama-ku, Kawasaki, Kanagawa 214-8571, Japan
\texttt{shin.takasuka@gmail.com, mukai@meiji.ac.jp}}
\date{}

\begin{document}

\maketitle
\doublespacing

\begin{abstract}
This study aims to enhance the quality of music generation using Transformers by incorporating meta-information.
While Transformer-based approaches are effective at capturing long-term dependencies in musical compositions, the music they generate often suffers from issues such as excessive repetition or duplication of notes, leading to unnatural melodies. To address these limitations, we propose Musical Attention, a mechanism that incorporates meta-information such as bar numbers, key, signatures, and tempos into the attention process.
Musical Attention explicitly leverages both the structural properties of music and its associated metadata, enabling the Transformer's attention mechanism to operate more effectively and thereby improving the quality of the generated output.
In our framework, each musical note is represented as a combination of five events-pitch, bar number, onset, duration, and velocity in addition to the three metadata elements. The attention mechanism is then modified to reflect the correlations among these eight features, allowing the model to better capture the inherent characteristics of musical composition.
Experimental results demonstrate that the model incorporating Musical Attention outperforms prior methods, such as Full Attention and Strided Attention, in terms of musical coherence, variation, and overall quality.
Notably, it significantly reduces repetition and enhances the model's ability to generate diverse, harmonically consistent melodies. 
Musical Attention thus represents a meaningful advancement in AI-driven music generation, facilitating the creation of more natural and expressive compositions. 
\end{abstract}


\section{Introduction}\label{sec:intro}

With the rapid development of machine learning, music generation \cite{musicvae, midivae, cyclegan-music} has made significant progress. 
In particular, Transformer-based approaches \cite{transformer} have emerged as promising techniques. 
Transformers have achieved remarkable results in natural language processing, making them readily applicable to music generation. 
However, the quality of music generated by Transformer models \cite{music-bert, midi-bert, musicgen} still has considerable room for improvement. 

Specifically, issues such as inconsistency in melody, lack of emotional expression, and deficiencies in musical flow and diversity highlight the need for further research and enhancements. 
In the current study, we attempted to improve the quality of generated music using Transformers. 
Transformers are known to rely on absolute position representations, utilizing positional sinusoidal waves for each positional input encoding or learned positional encoding. 
Musical compositions consist of multiple elements, such as pitch, duration, and velocity, where relative differences between them are more important than absolute placements. 

To better capture these relationships between musical elements, Shaw et al. \cite{relative-attention} introduced relative attention, which represents the distance between self-attention mechanisms at two positions. 
Relative attention requires $O(L^{2}D)$ memory, where $L$ is the sequence length and $D$ is the dimension of hidden states in the model. 
Music Transformer has reduced the required memory to $O(LD)$, demonstrating the practical use of relative attention for long sequences. 

In a previous study \cite{music-transformer}, the beginning part of a musical composition was used as input data to generate the successive part. 
In those generative experiments, simultaneous notes with the same pitch or repeated phrases were frequently generated due to the probabilistic nature of musical note generation. 
In the current study, we aimed to generate high-quality and consistent musical pieces by improving the Music Transformer.

Several studies using Transformers generate musical pieces from text \cite{musiclm, mulan}. 
Music generative models based on text are suitable for producing natural-sounding compositions; however, they sometimes generate unnatural music depending on the input text. 
This is partly because natural language fails to fully express the precise content of musical theory when text data is used for training. 
Additionally, the interpretation of music heavily relies on the interpreter's sense and/or musical knowledge, leading to individual differences in the fidelity of generated music to textual descriptions. 

To address these issues, the current study employed meta-information about musical compositions, such as the number of bars, key, and tempo, as inputs. 
We propose a music generation model that adheres to these meta-information parameters to ensure structural consistency.

The remainder of this paper is organized as follows. 
Section \ref{sec:method} provides an exposition of the adjusted Music Transformer based on musical causal relationships. 
Section \ref{sec:results} presents the experimental results and evaluations. 
Finally, Section \ref{sec:conclusion} concludes the paper and discusses future work.

\section{Related Work}\label{sec:related_work}

Traditionally, two major approaches have been employed to encode symbolic music: piano-roll representations and MIDI-based formats.  
Piano-roll-based methods, such as MIDINet \cite{midinet} and MuseGAN \cite{musegan}, 
treat MIDI data as two-dimensional images, with time steps and pitches forming the axes. 
These representations are typically processed using convolutional neural networks (CNNs) or generative adversarial networks (GANs) \cite{gan}. 
Such approaches have achieved notable success by capturing complex temporal dependencies and improving the coherence and diversity of generated musical patterns.

In contrast, language-model-based approaches represent MIDI sequences as one-dimensional token streams based on events such as note-on, note-off, and time-shift. 
Models such as MusicBERT \cite{bert} and MuseNet \cite{musenet} adopt this paradigm. 
The Music Transformer \cite{music-transformer}, for instance, utilizes only the decoder portion of the Transformer architecture and conditions generation on the initial segment of a MIDI sequence. 
Since attention is restricted to preceding tokens, the model naturally tends to generate repetitive structures. To mitigate the high memory requirements of standard attention, the Skewing algorithm \cite{music-transformer} was proposed to reduce computational overhead.

Another stream of research focuses on enhancing the preprocessing of symbolic music data. 
In the REMI framework \cite{remi}, generation quality was improved through a detailed redesign of the event vocabulary. 
By jointly encoding note-on events and their corresponding durations, 
REMI addressed common problems such as notes failing to turn off correctly. 
MuseMorphose \cite{musemorphose} introduced bar-level control over rhythm intensity—measured by the ratio of onsets to beats—and polyphony score—defined as the average number of notes played within a quarter-note duration—enabling finer-grained control over generative outputs.

The Sparse Transformer \cite{sparse-transformer}, later adopted in models like GPT-3 \cite{gpt3} and InstructGPT \cite{instructgpt}, introduced sparsity into the attention mechanism to reduce memory usage and computational cost. 
It also proposed two novel attention patterns: Strided Attention, which is particularly effective for data with cyclic structures such as music or images, 
and Fixed Attention, which is more suitable for handling sequential text data.

More recently, MusicLM \cite{musiclm} has demonstrated the ability to generate high-quality music from textual prompts. 
This model integrates AudioLM \cite{audiolm}—which combines SoundStream \cite{soundstream} and wav2vec-based models such as Wav2Vec2 or Wav2BERT \cite{wav2bert}—and MuLan \cite{mulan}, 
a model that maps both audio and text into a shared latent space. 
While such text-conditioned models are capable of producing coherent and expressive music, 
they may sometimes fail to reflect the intended musical structure due to the inherent ambiguity and limitations of natural-language descriptions.

In response to these challenges, our study proposes the Music Attention mechanism, 
a Transformer-based architecture that incorporates meta-information such as bar numbers, key signatures, and tempo. 
This approach is designed to capture musical causal relationships and improve the structural consistency and expressive quality of the generated music.

\section{Technical Background}\label{sec:tech_background}

In the Transformer architecture, input sequences are first embedded as $X = \{x_{1}, \ldots, x_{T}\}$, where each $x_{t} \in \mathbb{R}^{d}$.  
Here, $d$ denotes the embedding dimension of the model.

The Transformer comprises multi-head attention (MHA) and a feedforward network (FFN). Each layer applies residual connections and layer normalization as follows:
\begin{align}
    s^{l}_{t} &= h^{l-1}_{t} + \text{MHA}(\text{LayerNorm}(h^{l-1}_{t} \mid H^{l-1}_{t})) \\
    h^{l}_{t} &= s^{l}_{t} + \text{FFN}(\text{LayerNorm}(s^{l}_{t}))
\end{align}
Here, $H^{l-1} = [h^{l-1}_{1}, \ldots, h^{l-1}_{T}] \in \mathbb{R}^{T \times d}$, and $s^{l}_{t}, h^{l}_{t} \in \mathbb{R}^{d}$.

The MHA component computes queries, keys, and values:
\begin{align}
    Q^{l} = H^{l-1}W^{l}_{Q}, \quad K^{l} = H^{l-1}W^{l}_{K}, \quad V^{l} = H^{l-1}W^{l}_{V}
\end{align}
where $W^{l}_{Q}, W^{l}_{K}, W^{l}_{V} \in \mathbb{R}^{d \times d}$ are trainable projection matrices.

The query matrix $Q^{l}$ is split into $M$ heads: $[Q^{l,1}, \ldots, Q^{l,M}]$, with $Q^{l,m} \in \mathbb{R}^{T \times \frac{d}{M}}$ for all $m \in \{1, \ldots, M\}$.  
Similarly, $K^{l}$ and $V^{l}$ are also divided across the $M$ heads.

For each attention head, the attention mechanism is computed as follows:
\begin{align}
    \text{Att}^{l,m} = \text{softmax}\left(\frac{Q^{l,m}{K^{l,m}}^\top + S_{\text{rel}}}{\sqrt{d/M}}\right)V^{l,m}
\end{align}
where $\text{Att}^{l,m} \in \mathbb{R}^{T \times \frac{d}{M}}$, and $S_{\text{rel}}$ denotes the relative positional encodings introduced by Shaw et al. \cite{relative-attention}.  
These encodings allow the model to consider the relative positions of tokens rather than absolute ones.

Cheng et al. \cite{music-transformer} proposed the Skewing algorithm to reduce the memory complexity of relative attention from $O(L^2D)$ to $O(LD)$, making it more feasible for long sequences.  
The outputs from all attention heads are concatenated and linearly projected using a trainable weight matrix $W^{l}_{O} \in \mathbb{R}^{d \times d}$.

In this study, temperature sampling is employed to generate musical sequences.  
In Large Language Models (LLMs), temperature sampling is a technique used to adjust the sharpness of the output probability distribution.  
The softmax function is scaled by a temperature parameter $t$, as defined in Equation~\eqref{eq:temp}:
\begin{align}
    f(x_{i},t) = \frac{e^{x_{i}/t}}{ \sum^{K}_{j=1} e^{x_{j}/t}} \label{eq:temp}
\end{align}

\begin{figure}[tbp]
    \begin{minipage}[t]{0.495\linewidth}
        \centering
        \includegraphics[width=\linewidth]{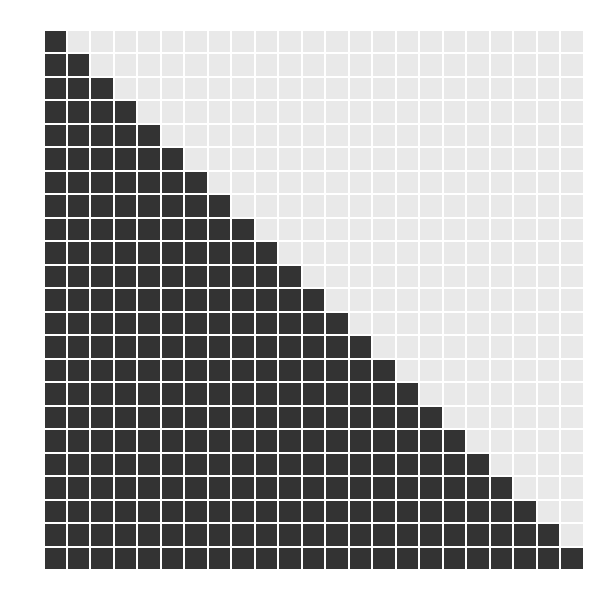}
        \subcaption{Full attention}
        \label{fig:plain}
    \end{minipage} 
    \begin{minipage}[t]{0.495\linewidth}
        \centering
        \includegraphics[width=\linewidth]{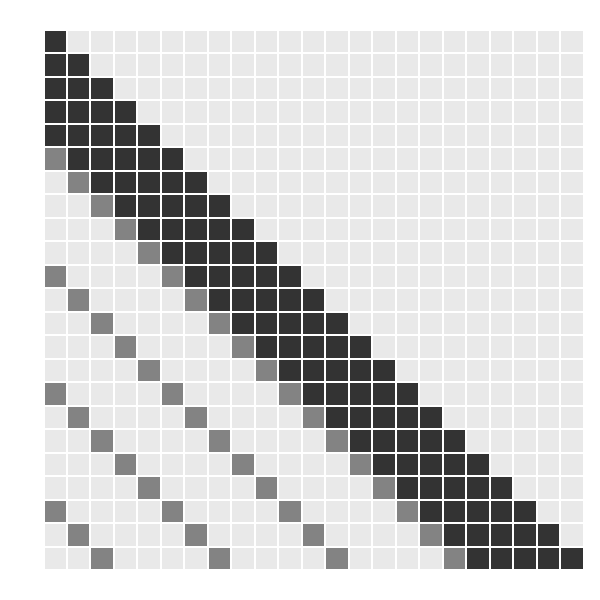}
        \subcaption{Strided attention}
        \label{fig:stride}
    \end{minipage} 
    \caption{Comparison of different attention mechanisms for multitrack music data.
    Left: Full attention, where each token attends to every position in the sequence.
    Right: Strided attention, which consists of two patterns: local attention to neighboring tokens capturing local structures, and sparse attention to distant tokens modeling long-range dependencies.}
    \label{fig:mask}
\end{figure}

The Sparse Transformer introduces sparsity into the MHA mechanism to reduce both computational cost and memory usage \cite{sparse-transformer}.  
Specifically, Strided Attention employs two patterns: one that periodically attends to distant positions, and another that focuses on neighboring tokens.  
This dual-pattern structure is especially beneficial for music, which often contains both local rhythmic structures and long-range dependencies.  
Figure~\ref{fig:mask} compares full attention and strided attention in the context of multitrack music data.

\section{Method}
\label{sec:method}

\subsection{Model Overview}

\begin{figure}[tbp]
    \centering
    \includegraphics[width=\linewidth]{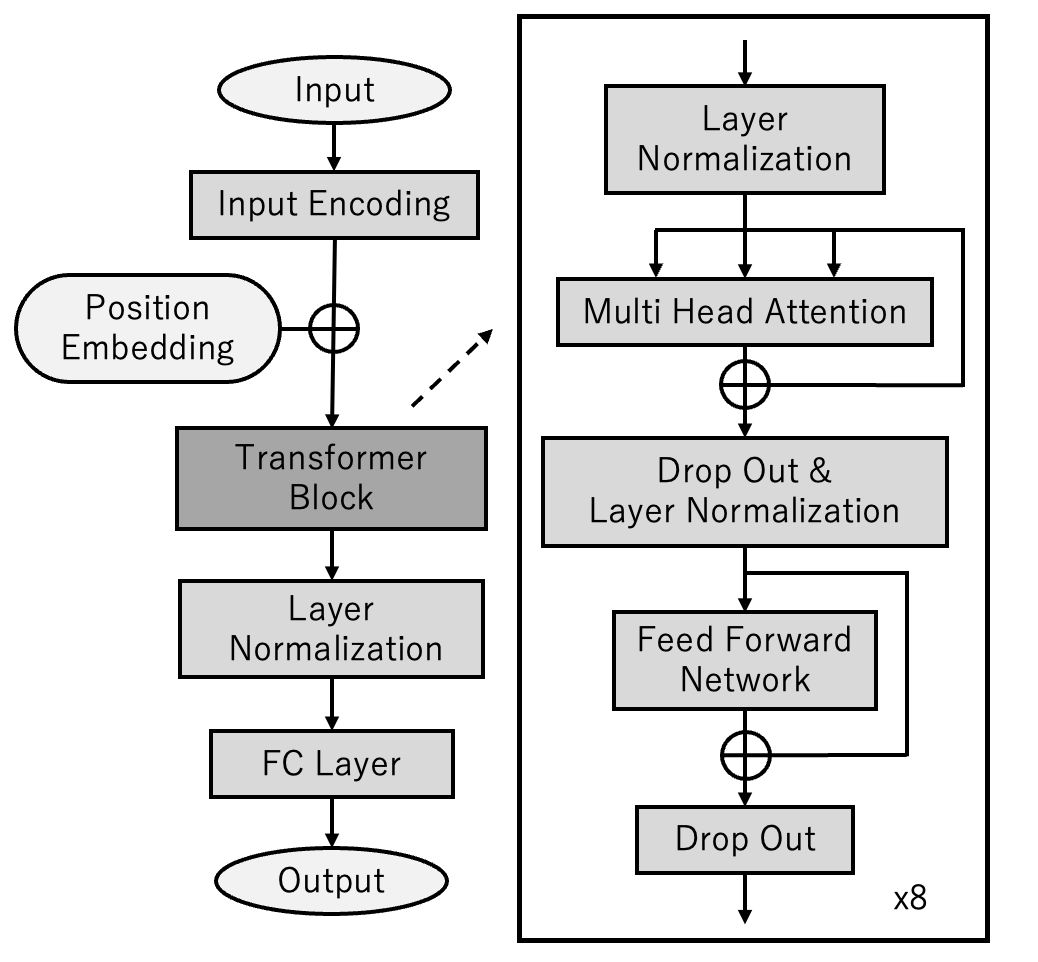}
    \caption{
    Model architecture of the Music Attention Transformer.
    The input sequence consists of a music event vector $N_j = [i_j, p_j, b_j, s_j, d_j, v_j]$ and meta information such as bar number $B$, key $K$, and tempo $T$.
    These vectors are concatenated with positional information and fed into the Transformer encoder.
    Through autoregressive training that predicts the next token from the previous one, the model learns contextual information and structural patterns in the music.} 
    \label{figure:model_structure}
\end{figure}

As shown in Figure~\ref{figure:model_structure}, the Music Attention Transformer pre-trains a Transformer encoder using masked language modeling (MLM),
where a portion of tokens in the input music sequence is masked and then predicted.  
To efficiently tokenize input sequences, each note is represented using six musical event types.  
The embeddings of these events are linearly combined into a single vector, which is then added to a positional embedding and provided as input to the Transformer encoder.

\subsection{Preprocessing of MIDI}

\begin{figure}[tbp]
    \centering
    \includegraphics[width=\linewidth]{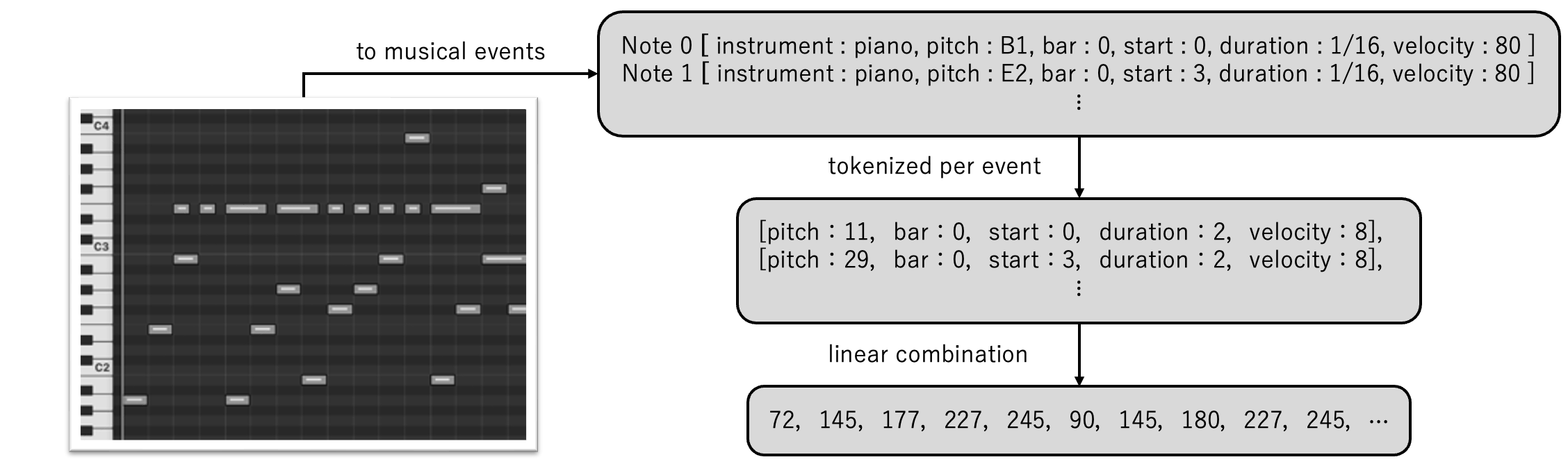}
    \caption{Overview of MIDI data preprocessing.
    First, MIDI files collected from the Lakh MIDI Dataset were analyzed using \cite{miditoolkit}.
    The drum tracks were removed, and note information was converted into six types of music event vectors $N_j = [i_j, p_j, b_j, s_j, d_j, v_j]$.
    Subsequently, meta information such as bar number $B$, key $K$, and tempo $T$ was extracted and concatenated with the note vectors to form the final input sequence $X = [B, K, T, N_1, \dots, N_n]$.}
    \label{figure:preprocess}
\end{figure}

Figure~\ref{figure:preprocess} provides an overview of the MIDI preprocessing pipeline used in this study.

A total of 45,129 MIDI files were used for training.  
From these files, 250,545 single-track sequences and 347,309 multi-track sequences were extracted.  
The data were collected from the Lakh MIDI Dataset \cite{dataset} and processed using \cite{miditoolkit} to convert each file into six types of musical event vectors.  
Each sequence was then divided into segments with a maximum length of $T$ and used as training data.

Each note was encoded as a six-dimensional vector $N_{j} = [i_{j}, p_{j}, b_{j}, s_{j}, d_{j}, v_{j}]$, where:  
$i$ is the instrument number,  
$p$ is the pitch,  
$b$ is the bar number,  
$s$ is the starting position,  
$d$ is the duration,  
and $v$ is the velocity.

We excluded drum tracks and used ten instrument classes for $i$.  
For pitch $p$, we used 84 MIDI note values spanning C0 to B6.  
Bar number $b$ represents the position within the piece, while starting position $s$ indicates the temporal offset within the bar, quantized into 48 subdivisions (i.e., 0 to 47).  
Note duration $d$ was encoded as two discrete values representing durations from a 48th note to a whole note, normalized into a range of 0 to 11.  
Velocity $v$ was mapped from a MIDI range of 40 to 115 into 16 discrete levels (0–15).

We also extracted meta-information:  
$B$ (the total number of bars),  
$K$ (key signature, estimated from the full sequence and spanning C major to B major, including relative minors),  
and $T$ (tempo, quantized from 50–200 BPM into 16 levels).  
The final input sequence $X$ consists of the concatenated meta-information $[B, K, T]$ and note event vectors $[N_{1}, \ldots, N_{n}]$.

\subsection{Musical Attention}\label{sec:musical-attention}

\begin{figure}[tbp]
    \centering
    \includegraphics[width=\linewidth]{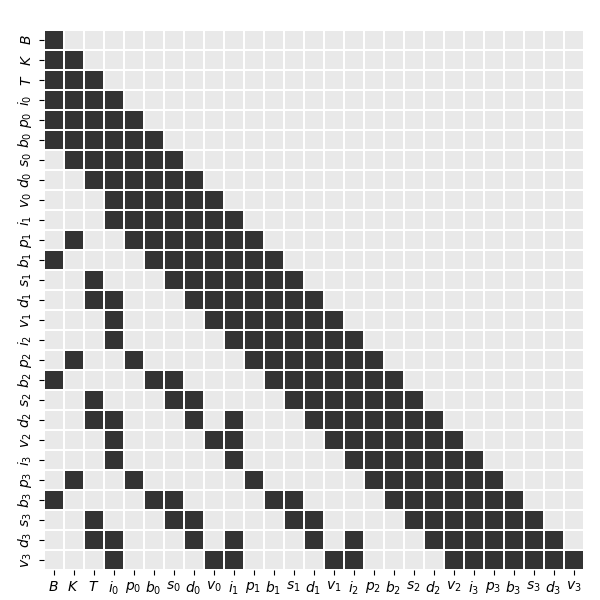}
    \caption{Mechanism of Musical Attention.
    Two additional attention patterns are introduced on top of standard self-attention:
    (a) attention focused on a limited number of preceding tokens, and
    (b) attention guided by musical dependencies.
    This design allows the model to naturally incorporate music-theoretical constraints,
    such as pitch dependency on key, rhythmic structures based on bar positions and onset timing,
    and dynamic variations caused by different instrument types.}
    \label{figure:musical-attention}
\end{figure}

We propose Musical Attention, a novel attention mechanism designed to generate music conditioned on meta-information such as bar count, key, and tempo.  
Figure~\ref{figure:musical-attention} illustrates the attention structure used for multi-track music generation.

Musical Attention modifies the standard self-attention mechanism to incorporate causal and structural dependencies in music.  
It uses two main attention patterns:  
(1) attending to a limited set of preceding tokens within the same musical context, and  
(2) referencing specific tokens that share the same attribute (e.g., same pitch or bar) between the current and previous positions.

This mechanism explicitly integrates several music-theoretic constraints:

\begin{itemize}
    \item Pitches $p_j$ are contextually dependent on the key signature $K$.
    \item The bar number $b_j$ is constrained by the global bar count $B$ and all previous starting positions $s_k$.
    \item The starting position $s_j$ is influenced by the tempo $T$ and the durations $d_k$ of nearby notes.
    \item The note duration $d_j$ increases with faster tempos and is influenced by the instrument type $i_k$.
    \item The velocity $v_j$ varies by instrument type and performance context.
\end{itemize}

Here, $j$ denotes the current token position, while $k$ indexes a referenced token (e.g., a preceding or contextually related note) used to model dependencies.

We compare Musical Attention with two baseline mechanisms: Full Attention (used in Music Transformer) and Strided Attention (used in Sparse Transformer).

\section{Experiments}\label{sec:experiments}

\begin{figure}[tbp]
    \centering
    \includegraphics[width=\linewidth]{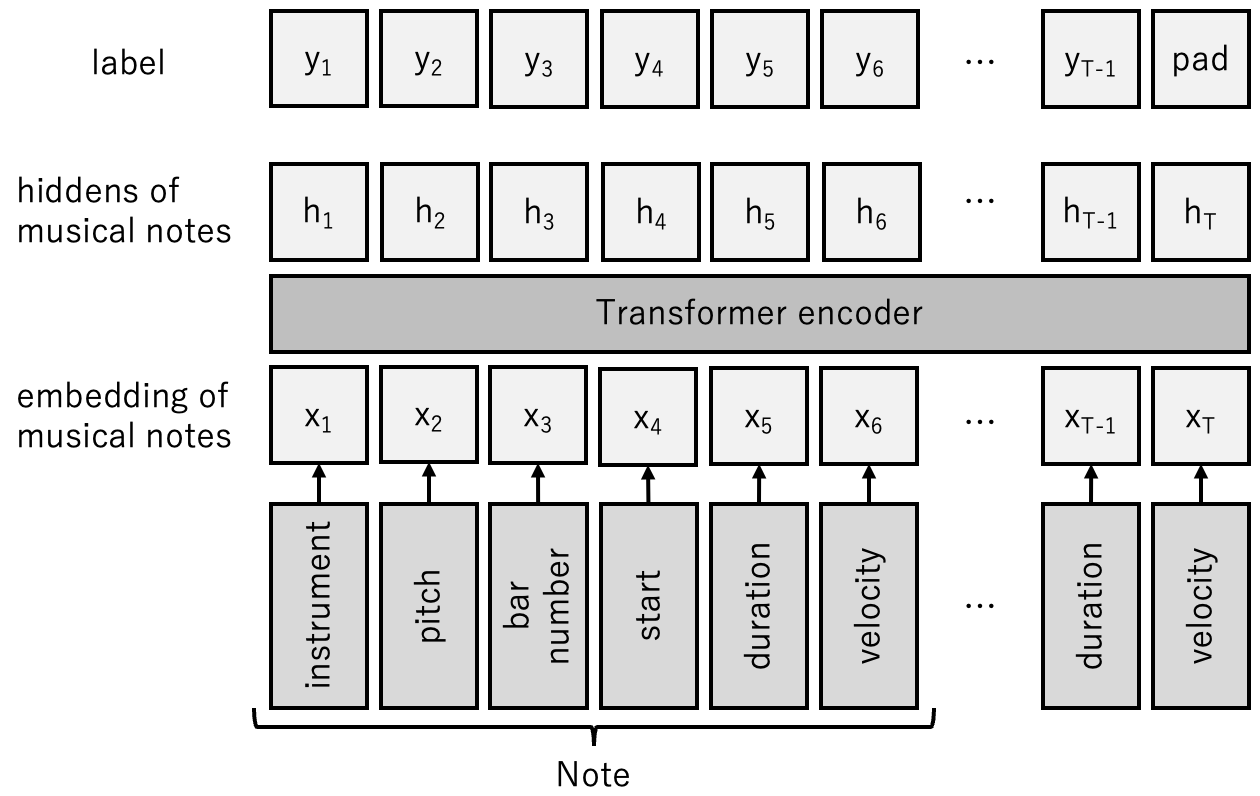}
    \caption{
    The training process of the Transformer model. 
    MIDI data are preprocessed and input to the encoder. 
    Here, $x_t \in \mathbb{R}^{d}$ is an embedding, and $y_t \in \mathbb{R}^{d}$ is the one-hot encoding of the $(t-1)$-th event in the MIDI sequence.
    The model learns the structural patterns of the music by predicting the next event from the token sequence presented immediately before.
    }
    \label{fig:model_learning}
\end{figure}

MIDI data are vectorized through preprocessing and subsequently fed into the encoder of the Transformer for training. 
Figure~\ref{fig:model_learning} illustrates the training process. 
Here, $x_{t} \in \mathbb{R}^{d}$ denotes the embedded input, and $y_{t} \in \mathbb{R}^{d}$ represents the one-hot encoding of the $(t-1)$-th event in the MIDI sequence. 

To quantitatively evaluate the generated musical compositions, we employed the following four metrics.

\subsection{Evaluation Metrics}

\subsubsection{Token Error}
Each note is encoded using six tokens, e.g., $[i, p, b, s, d, v]$. 
If the order of these tokens is disrupted, the MIDI data cannot be reconstructed properly. 
The number of incorrectly generated tokens is defined as the Token Error.

\subsubsection{Note Error}
Note Error is defined as the total number of the following three types of errors occurring at the $j$-th note:
\begin{itemize}
    \item \textit{Simultaneous same note error}: $i_{j} = i_{j+1} \,\land\, p_{j} = p_{j+1} \,\land\, b_{j} = b_{j+1} \,\land\, s_{j} = s_{j+1}$
    \item \textit{Bar number error}: $b_{j} > b_{j+1}$
    \item \textit{Start time error}: $b_{j} = b_{j+1} \,\land\, s_{j} > s_{j+1}$
\end{itemize}
A simultaneous same note error occurs when the pitch, bar number, and start time of the $j$-th and $(j+1)$-th notes are identical.  
A bar number error occurs when the bar number of the $j$-th note exceeds that of the $(j+1)$-th note.  
A start time error occurs when, within the same bar, the $j$-th note begins after the $(j+1)$-th note.  
Collectively, these metrics indicate whether the model failed to maintain proper temporal order in the note sequence.

\subsubsection{Bar Error}
Bar Error is defined as the difference in the number of bars between the input data and the generated music.  
The number of bars in the generated composition is determined using the last bar number, and discrepancies are measured in units of 48th notes.

\subsubsection{Key Error}
Key Error represents the proportion of notes in the generated music that do not belong to the key of the input data.

We evaluated the generated music using model accuracy as well as the median and mean values of these four musical evaluation metrics.

\subsection{Experimental Setup}
We compared different attention mechanisms for single-track music generation.
We used piano-only single-track data for preprocessing, and the training sequences were divided into segments of length 2048.
We trained three models: Full Attention, Strided Attention, and the proposed Musical Attention.
Each model consisted of 8 Transformer layers with an embedding size of 512 and 8 attention heads.
The feed-forward network (FFN) had hidden layer sizes of 512 and 256.
The batch size was set to 8, and training was conducted for a total of 207,164 steps.
Each model generated 100 musical compositions under fixed input conditions: 16 bars in the key of C major and a tempo of 80 BPM.

Next, we evaluated the generation of multi-track music.  
As in the single-track experiments, we tested the three attention mechanisms.  
Since multi-track music typically contains fewer notes per track, the number of input bars was reduced to 12.

We also examined the effect of temperature on music generation.  
This experiment used the Musical Attention model trained on single-track data.  
Input conditions were identical to those in the single-track experiment.

\section{Results}\label{sec:results}

Table~\ref{table:exp1} presents the results for single-track music generation.  
The Full Attention model achieved the highest training accuracy at 78.12\%.  
It also recorded the highest average Token Error (0.03).  
Strided Attention performed best in terms of Note Error, with an average of 0.18.  
Musical Attention outperformed the other models in both Bar Error and Key Error, achieving the lowest averages of 0.86 and 1.97, and medians of 1.19 and 1.47, respectively.

\begin{table}[tbp]
    \centering
    \caption{Results in Single-track Music Generation}
    \label{table:exp1}
    \small
    \begin{tabular}{lc|ccc}
        \hline
        \multicolumn{2}{l|}{Metric} & \multicolumn{3}{c}{Attention} \\
        \multicolumn{2}{l|}{} & Full (base) & Strided & Musical \\
        \hline
        \multicolumn{2}{l|}{Accuracy (\%)} & \textbf{78.12} & 77.98 & 78.07 \\
        Token & mean & \textbf{0.03} & 0.10 & 0.04 \\
              & median & 0 & 0 & 0 \\
        Note  & mean & 0.42 & \textbf{0.18} & 0.25 \\
              & median & 0 & 0 & 0 \\
        Bar   & mean & 1.21 & 1.12 & \textbf{0.86} \\
              & median & 1.26 & 1.28 & \textbf{1.19} \\
        Key   & mean & 4.69 & 5.07 & \textbf{1.97} \\
              & median & 2.69 & 4.40 & \textbf{1.47} \\
        \hline
    \end{tabular}
\end{table}

Table~\ref{table:exp2} shows the results for multi-track music generation.
The Strided Attention model achieved the highest training accuracy at 81.17\%.  
Musical Attention recorded the lowest Token Error, with an average of 0.04.  
Strided Attention performed best on Note Error, achieving an average of 0.29.  
Musical Attention again showed the best performance in Bar and Key Error, with averages of 0.62 and 2.55, and medians of 0.00 and 1.18, respectively.

\begin{table}[tbp]
    \centering
    \caption{Results in Multi-track Music Generation}
    \label{table:exp2}
    \small
    \begin{tabular}{lc|ccc}
        \hline
        \multicolumn{2}{l|}{Metric} & \multicolumn{3}{c}{Attention} \\
        \multicolumn{2}{l|}{} & Full (base) & Strided & Musical \\
        \hline
        \multicolumn{2}{l|}{Accuracy (\%)} & 81.08 & \textbf{81.17} & 81.09 \\
        Token & mean & 0.08 & 0.07 & \textbf{0.04} \\
              & median & 0 & 0 & 0 \\
        Note  & mean & 0.50 & \textbf{0.29} & 0.50 \\
              & median & 0 & 0 & 0 \\
        Bar   & mean & 1.04 & 0.80 & \textbf{0.62} \\
              & median & 1.17 & 0.00 & 0.00 \\
        Key   & mean & 5.18 & 5.56 & \textbf{2.55} \\
              & median & 3.24 & 2.65 & \textbf{1.18} \\
        \hline
    \end{tabular}
\end{table}

Table~\ref{table:exp3} summarizes the results of the temperature sampling experiments.  
Both the average and median Token Error were 0.00 when the temperature was set to 0.5 or 0.75.  
The best Note Error and Bar Error results were obtained at a temperature of 1.0, with average values of 0.25 and 0.86, and medians of 0.00 and 1.19, respectively.  
For Key Error, the lowest average (0.10) and median (0.00) were recorded at a temperature of 0.5.

\begin{table}[tbp]
    \centering
    \caption{Results with Different Temperature Parameters}
    \label{table:exp3}
    \small
    \begin{tabular}{lc|ccccc}
        \hline
        \multicolumn{2}{l|}{Metric} & \multicolumn{5}{c}{Temperature ($t$)} \\
        \multicolumn{2}{l|}{} & 0.50 & 0.75 & 1.00 & 1.25 & 1.50 \\
        \hline
        Token & mean & 0.00 & 0.00 & 0.05 & 1.54 & 9.76\\
              & median & 0 & 0 & 0 & 1 & 10\\
        Note  & mean & 2.21 & 0.70 & \textbf{0.25} & 0.62 & 0.90\\
              & median & 0 & 0 & 0 & 0 & 1\\
        Bar   & mean & 4.49 & 2.52 & \textbf{0.86} & 1.20 & 4.81\\
              & median & 3.49 & 2.39 & \textbf{1.19} & 1.26 & 4.60\\
        Key   & mean & \textbf{0.10} & 0.73 & 1.97 & 18.89 & 28.85\\
              & median & 0.00 & 0.00 & 1.47 & 18.22 & 28.61 \\
        \hline
    \end{tabular}
\end{table}

\section{Discussion}\label{sec:discussion}

\subsection{Effectiveness of Musical Attention}

As shown in Table~\ref{table:exp1}, there was no significant difference in training accuracy among different attention mechanisms, indicating that even attention mechanisms with restricted scopes, such as Strided Attention and Musical Attention, can maintain learning performance.

Moreover, Musical Attention generated higher-quality music than both Full Attention (baseline) and Strided Attention.  
This improvement can be attributed to the mechanism where bar number tokens attend to the overall bar structure, allowing generation consistent with the desired number of bars.  
Similarly, pitch tokens attending to key tokens enables the generation of notes that remain within the scale of the input key.

Similar to the single-track experiment, Table~\ref{table:exp2} shows that restricting the scope of attention does not degrade training accuracy in the multi-track setting.  
Musical Attention again outperformed the other methods in generating music that accurately follows the intended bar structure and key, confirming the effectiveness of targeted attention mechanisms even in multi-track compositions.

As summarized in Table~\ref{table:exp3}, the Token Error decreased as the temperature parameter $t$ decreased.  
This phenomenon stems from the fact that higher temperature values flatten the softmax distribution, increasing the likelihood of generating inappropriate tokens.  

The optimal temperature for the Note Error was 1.0.  
At lower temperatures, the distribution becomes overly peaked, leading to reduced note diversity and more repetitive outputs.  
At higher temperatures, while diversity increases, so does the risk of generating inappropriate tokens, often resulting in repeated or incoherent outputs.  
Therefore, a temperature of 1.0 offered a balanced trade-off between diversity and correctness.

The Bar Error also achieved its lowest value at temperature 1.0, aligning with the Note Error.  
This suggests that increased note errors at extreme temperature values lead to inappropriate bar structures.

The Key Error was minimized at lower temperature settings, as the more peaked distributions reduce the likelihood of generating non-diatonic notes outside the input key.

Taken together, these findings suggest that a temperature setting of 1.0 is optimal for balancing accuracy and musical quality in generated compositions.

\subsection{Analysis and Limitations of Generated Music}

\begin{figure}[tbp]
    \begin{minipage}[t]{0.49\linewidth}
        \centering
        \includegraphics[width=\linewidth]{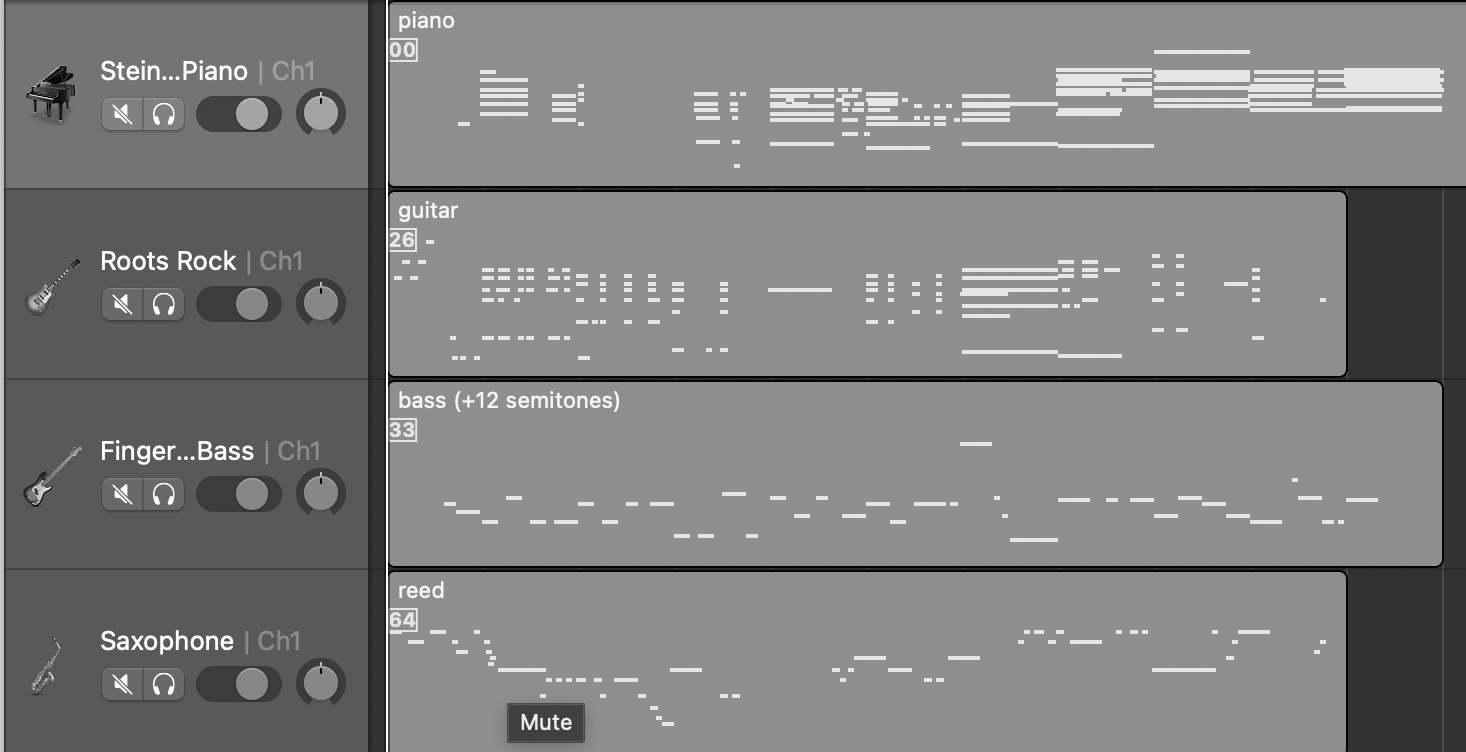}
        \subcaption{Full Attention}
        \label{fig:generate_full}
    \end{minipage}
    \hfill
    \begin{minipage}[t]{0.49\linewidth}
        \centering
        \includegraphics[width=\linewidth]{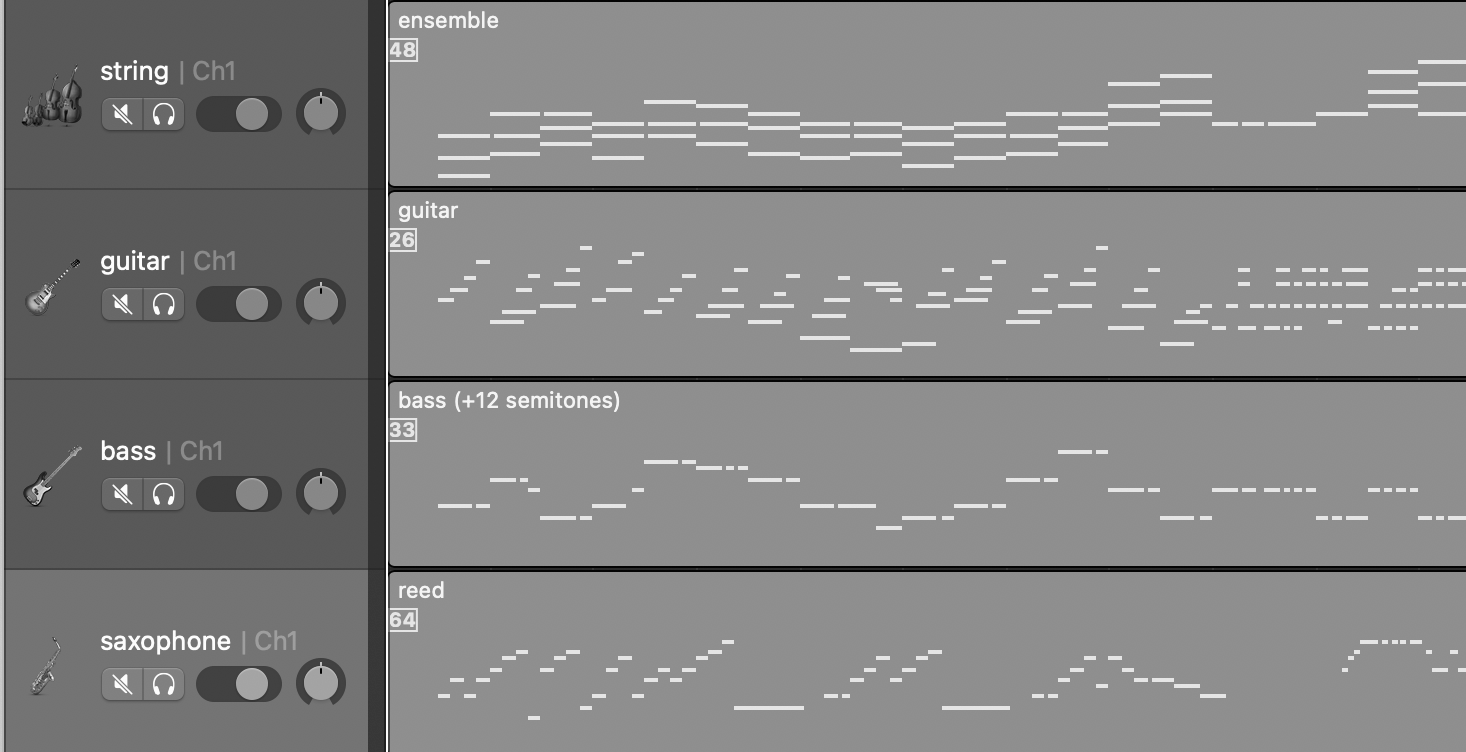}
        \subcaption{Musical Attention}
        \label{fig:generate_musical}
    \end{minipage}
    \centering
    \caption{
        Examples of multi-track music generated using the Full Attention and Musical Attention models.
        In the top guitar track, arpeggio phrases are generated that align with the underlying chord progression.
        The second track shows a bass line that follows the harmonic structure.
        In the third track, the string part consists of diatonic chords (e.g., Am, Dm, G, Am, F) within the key of C major, 
        while the bottom saxophone track presents a monophonic melody composed of notes in the same key.
        These results suggest that the model not only captures the overall musical structure but also learns instrument-specific performance styles.
    }
    \label{fig:generate}
\end{figure}

Figure~\ref{fig:generate} shows examples of multi-track compositions generated by the Full Attention and Musical Attention models.  
In the third track (strings), diatonic chords in the C major key—such as Am, Dm, G, Am, and F—are generated.  
The top guitar track contains arpeggiated phrases that follow these chords (see Appendices~\ref{sec:diatonic} and~\ref{sec:arpeggio} for definitions).  
The second track from the top (bass) follows the chord progression with a coherent bassline, while the bottom saxophone track features a melody composed entirely of single notes within the C major key.  

These results suggest that the model not only captures harmonic structure but also learns instrument-specific performance styles.  
For example, single-note melodies in the bass and guitar tracks indicate appropriate modeling of arpeggios and bass playing.  
This may be attributed to the attention mechanism from note duration tokens to instrument tokens, allowing the model to differentiate performance styles by instrument type.

\begin{figure}[tbp]
    \begin{minipage}[t]{0.49\linewidth}
        \centering
        \includegraphics[width=\linewidth]{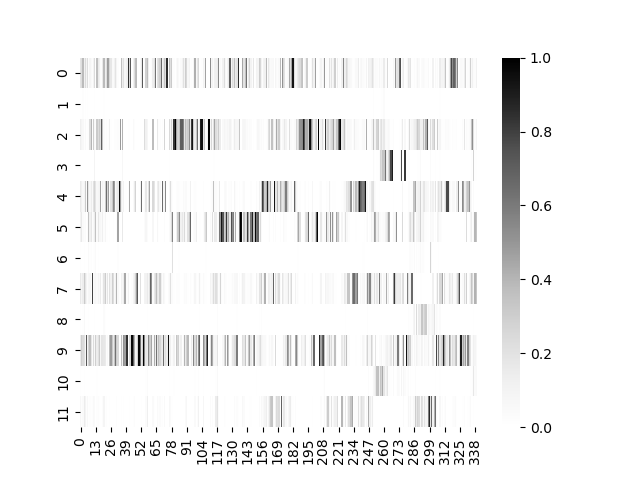}
        \subcaption{Full Attention}
        \label{fig:heatmap_full}
    \end{minipage}
    \hfill
    \begin{minipage}[t]{0.49\linewidth}
        \centering
        \includegraphics[width=\linewidth]{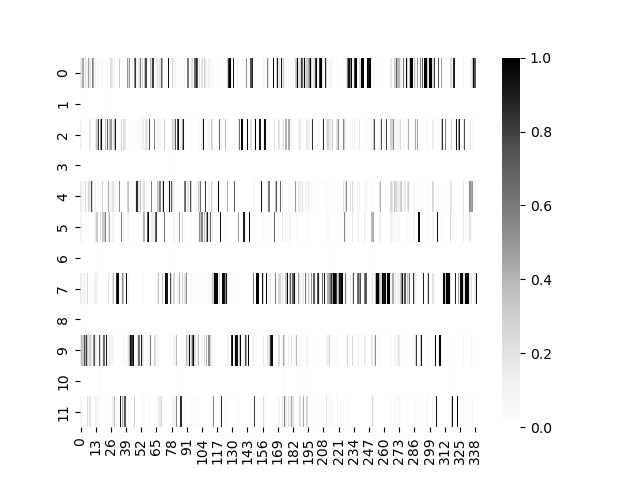}
        \subcaption{Musical Attention}
        \label{fig:heatmap_musical}
    \end{minipage}
    \centering
    \caption{
        Pitch heatmaps generated by the Full Attention and Musical Attention models.
        The figures visualize the generation probabilities of each pitch token, 
        where values from 0 to 11 correspond to the pitches C through B.
        The Musical Attention model exhibits a clearer probability distribution based on the input key and chord structure, 
        indicating generation more consistent with music-theoretical principles.
    }
\label{fig:pitch_heatmap}
\end{figure}

Figure~\ref{fig:pitch_heatmap} visualizes the generation probabilities of pitch tokens during sampling, 
where values from 0 to 11 correspond to pitches C through B.
Compared to the Full Attention model, the Musical Attention model exhibits a clearer concentration of probability mass on pitches within the input key, 
suggesting that the key information is more effectively reflected in pitch generation.

This study has two primary limitations.  
First, the generated music lacks dynamic variation; the volume remains constant across different instruments, failing to capture realistic dynamics.  
Second, while chord types align with the input key, the chord progressions sometimes appear unnatural or musically incoherent.  
Additional heatmaps and analyses related to these issues are provided in Appendix~\ref{sec:heatmap}.

\section{Conclusion}\label{sec:conclusion}

In this study, we proposed a music generation model incorporating meta-information such as bar numbers, key signatures, and tempo. 
We also developed Musical Attention, an updated self-attention mechanism specialized for music generation. 
Overall, we demonstrated the feasibility of generating musical compositions that adhere to these meta-information parameters. 
Furthermore, our quantitative evaluations showed that Musical Attention produces music of competitive quality compared to existing models.

One of the remaining challenges is that the current method does not fully utilize chord information or modulation within the compositions. 
By incorporating detailed information about chords and chord progressions, it will be possible to develop music generation systems capable of producing higher-quality musical pieces. 

Additionally, we did not label the generated samples with the specific attention conditions. 
By combining input information across multiple instruments, future work will enable the generation of compositions with specified instrumentations.

Music has traditionally been created based on music theory; however, the originality of a piece is often devised by deliberately deviating from established theory. 
It remains generally difficult for previous Transformer-based models to capture the inherent originality of generated musical pieces. 
While these models partially reflect music theory by extracting tendencies from vast training datasets, they lack explicit theoretical grounding. 

Preexisting text-based music generation models often suffer from text dependency, partly because input texts do not precisely reflect the nuances of music theory. 
By establishing a precise connection between natural language and musical theory within the latent space, the theoretical music generation approach presented in this study will be applicable to previously less-explored musical domains, such as Japanese popular music and other regional styles. 

\section*{Data Availability Statement}
The original datasets analyzed in this study are available at https://github.com/craffel/midi-dataset. The datasets generated during the current study are available from the corresponding author upon reasonable request.

\newpage
\onecolumn
\appendix
\section{Learning Curves} 
\subsection{Experiments for the Generation of Single-Track Music}\label{subsec:lc_exp1-2}

\begin{figure}[h]
    \begin{minipage}{0.495\textwidth}
        \centering
        \includegraphics[width=\linewidth]{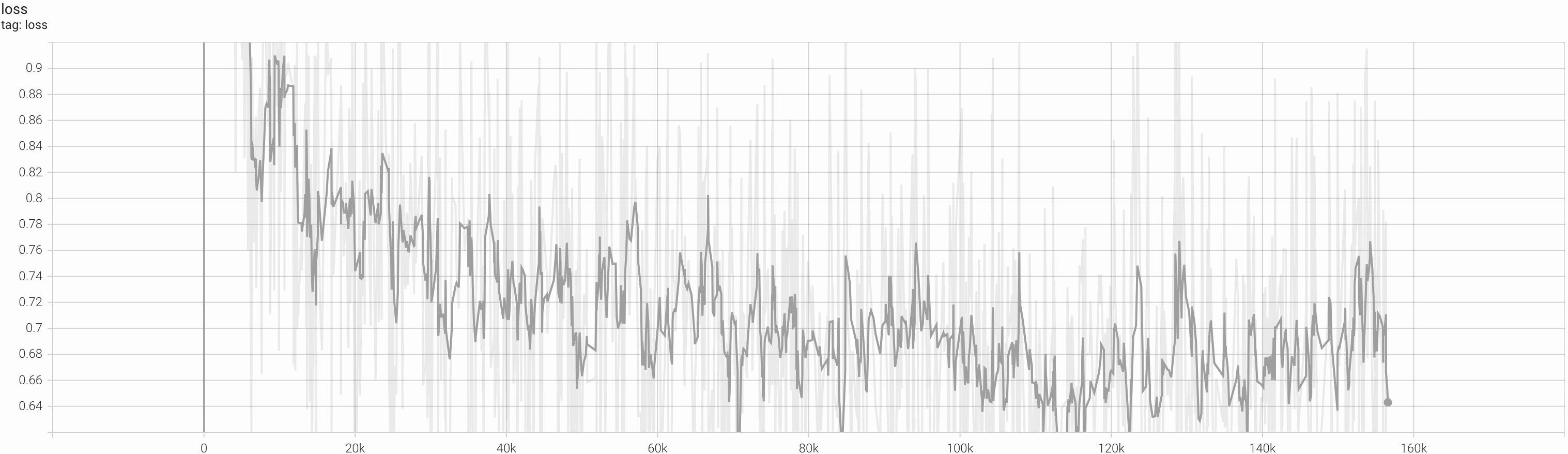}
        \subcaption{Train loss}
    \end{minipage}
    \begin{minipage}{0.495\textwidth}
        \centering
        \includegraphics[width=\linewidth]{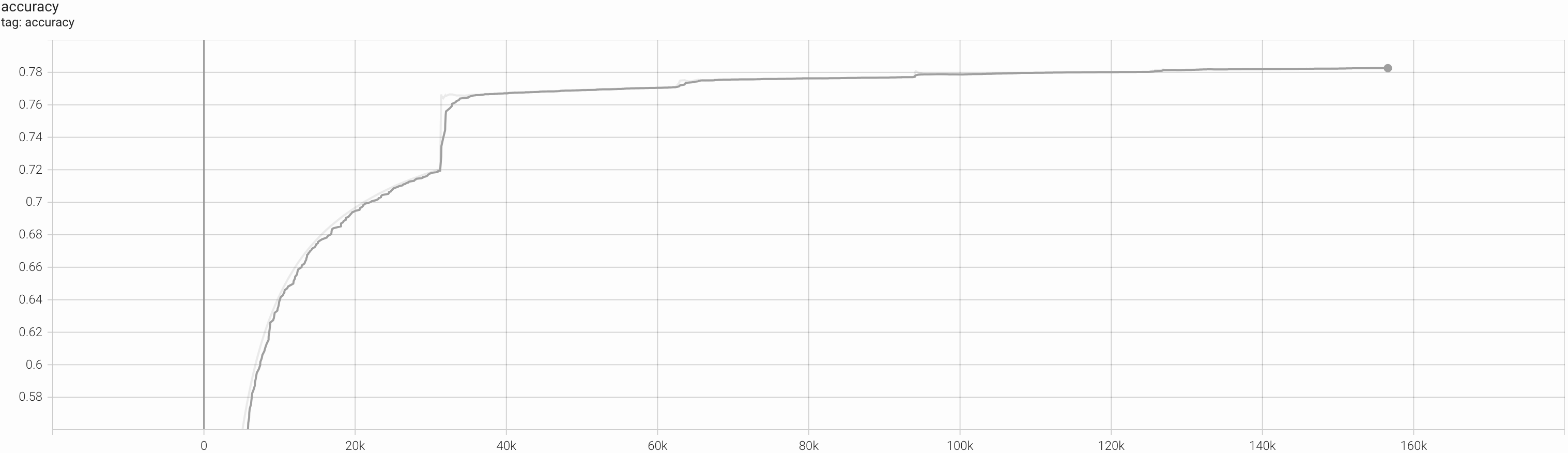}
        \subcaption{Train accuracy}
    \end{minipage}
    \begin{minipage}{0.495\textwidth}
        \centering
        \includegraphics[width=\linewidth]{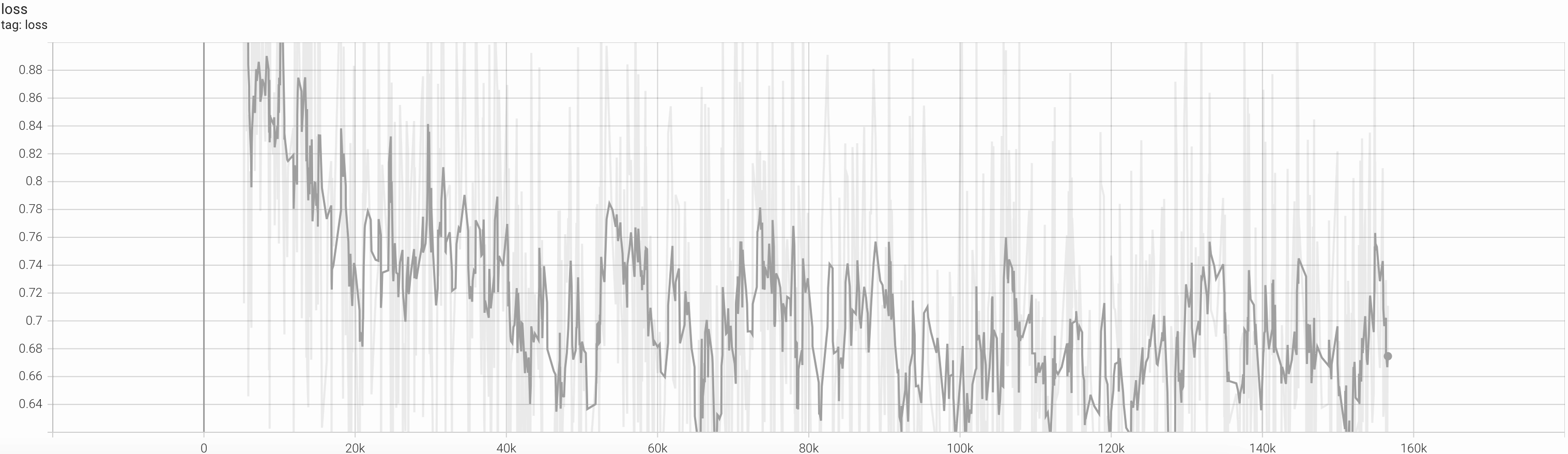}
        \subcaption{Eval loss}
    \end{minipage}
    \begin{minipage}{0.495\textwidth}
        \centering
        \includegraphics[width=\linewidth]{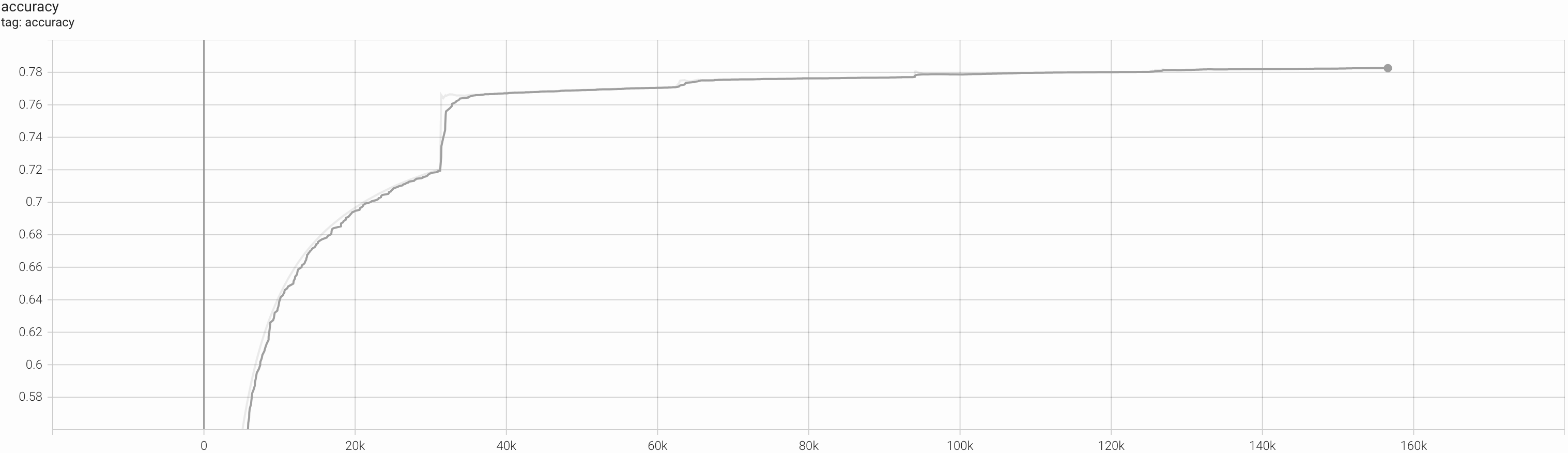}
        \subcaption{Eval accuracy}
    \end{minipage}
    \caption{Learning curves of single-track music generation.}
    \label{fig:single_learning_curve}
\end{figure}

Figure~\ref{fig:single_learning_curve} shows the learning curves for single-track music generation.

\subsection{Experiments for the Generation of Multi-Track Music} \label{sec:lc_exp2}

\begin{figure}[h]
    \begin{minipage}{0.495\textwidth}
        \centering
        \includegraphics[width=\linewidth]{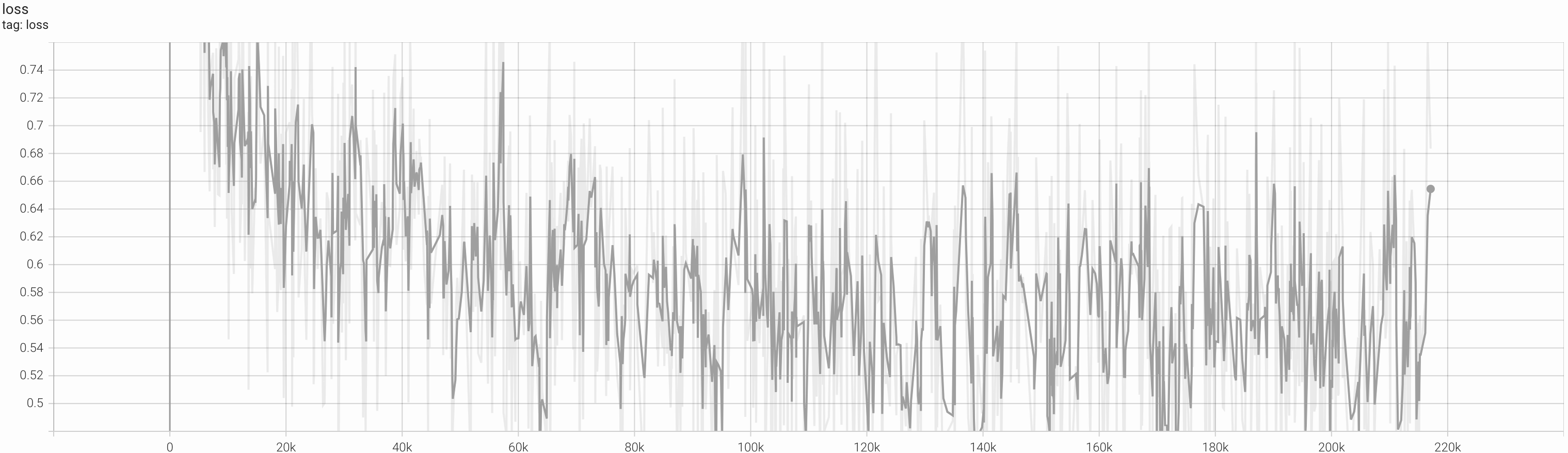}
        \subcaption{Train loss}
    \end{minipage}
    \begin{minipage}{0.495\textwidth}
        \centering
        \includegraphics[width=\linewidth]{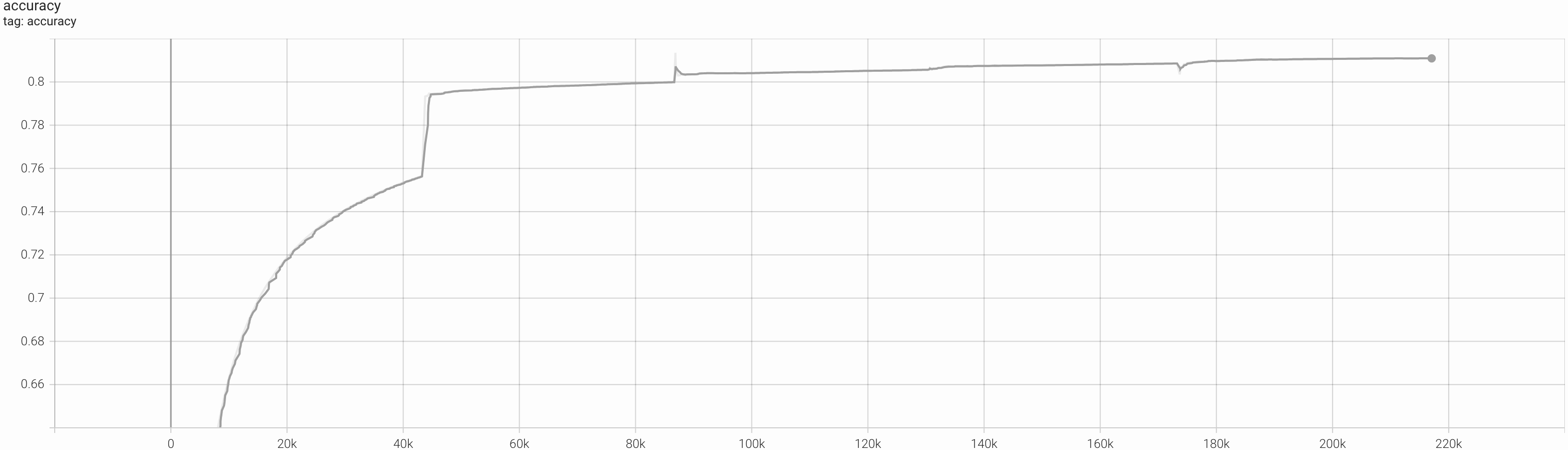}
        \subcaption{Train accuracy}
    \end{minipage}
    \begin{minipage}{0.495\textwidth}
        \centering
        \includegraphics[width=\linewidth]{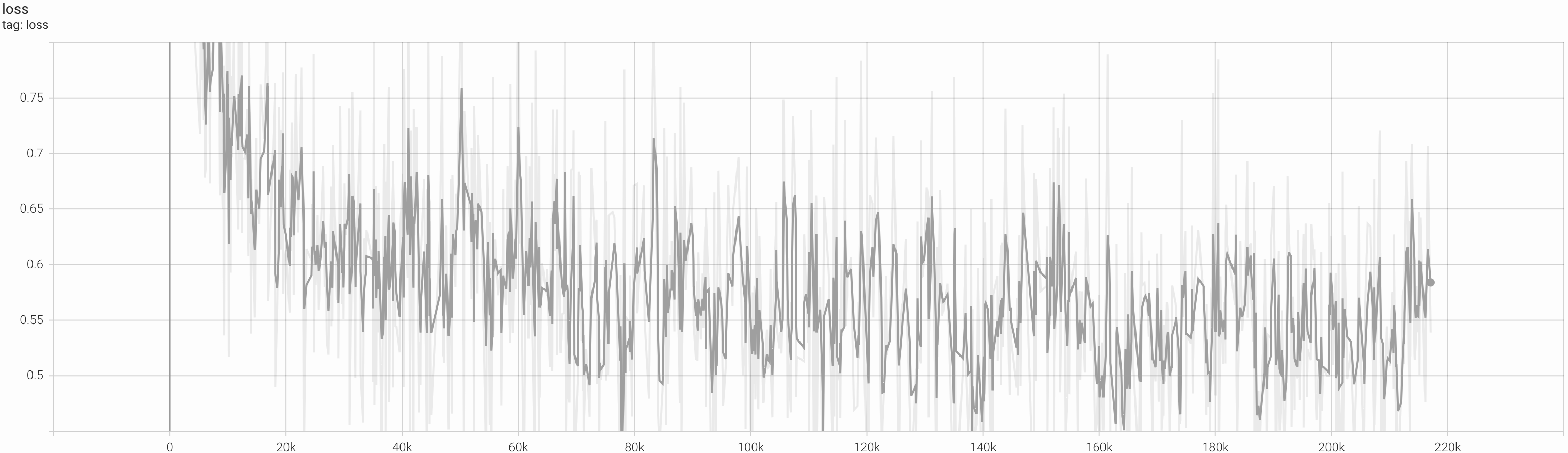}
        \subcaption{Eval loss}
    \end{minipage}
    \begin{minipage}{0.495\textwidth}
        \centering
        \includegraphics[width=\linewidth]{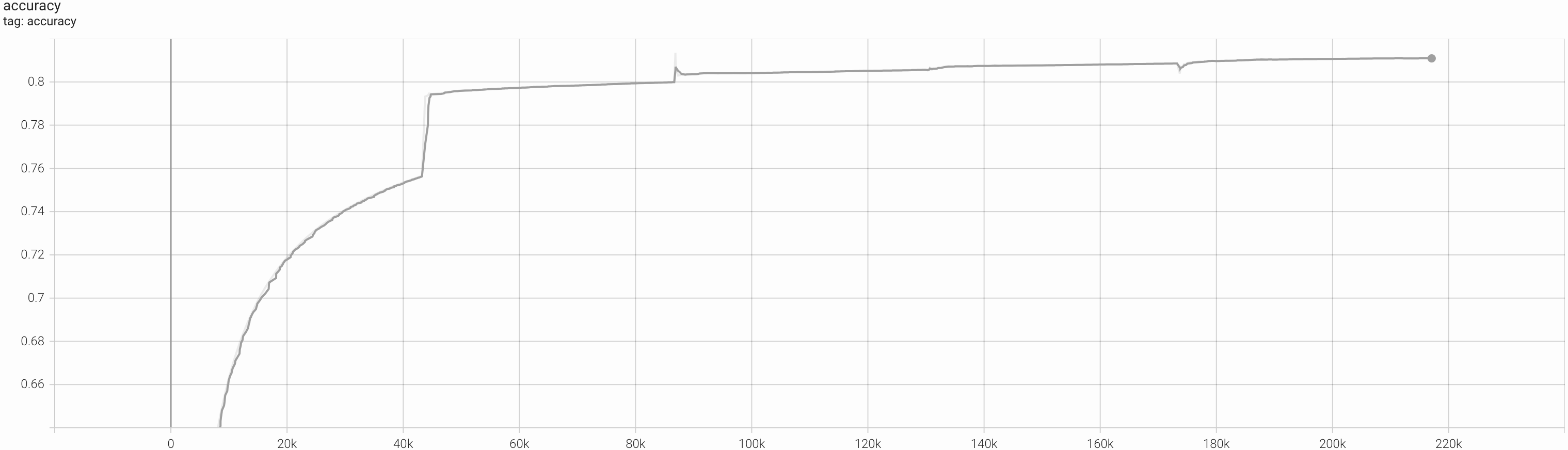}
        \subcaption{Eval accuracy}
    \end{minipage}
    \caption{Learning curves of multi-track music generation.}
    \label{fig:multi_learning_curve}
\end{figure}

Figure~\ref{fig:multi_learning_curve} shows the learning curves for multi-track music generation.

\newpage

\section{Heatmap During Music Generation} \label{sec:heatmap}
\begin{figure}[htbp]
    \begin{minipage}[t]{0.32\linewidth}
        \centering
        \includegraphics[width=\linewidth]{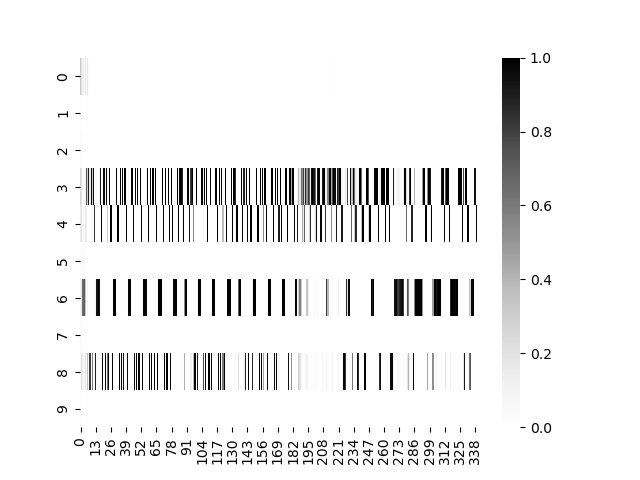}
        \subcaption{Instrument number}
        \label{fig:inst_heat_music}
    \end{minipage}\hfill
    \begin{minipage}[t]{0.32\linewidth}
        \centering
        \includegraphics[width=\linewidth]{fig/heatmap_music/pitch_heatmap.png}
        \subcaption{Pitch}
        \label{fig:pitch_heat_music}
    \end{minipage}\hfill
    \begin{minipage}[t]{0.32\linewidth}
        \centering
        \includegraphics[width=\linewidth]{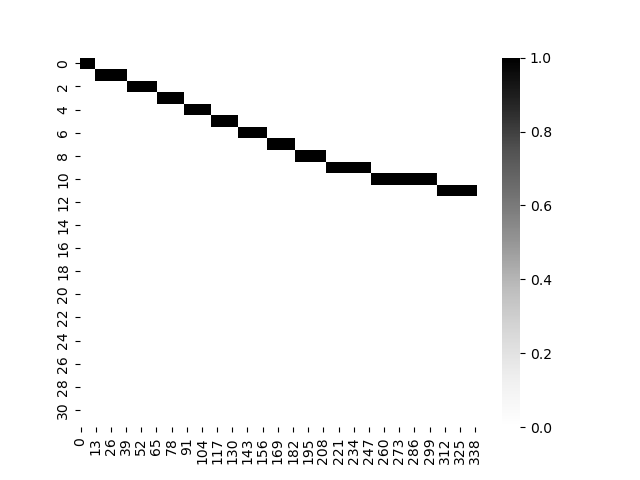}
        \subcaption{Bar number}
        \label{fig:bar_heat_music}
    \end{minipage}

    \vspace{1em} 

    \begin{minipage}[t]{0.32\linewidth}
        \centering
        \includegraphics[width=\linewidth]{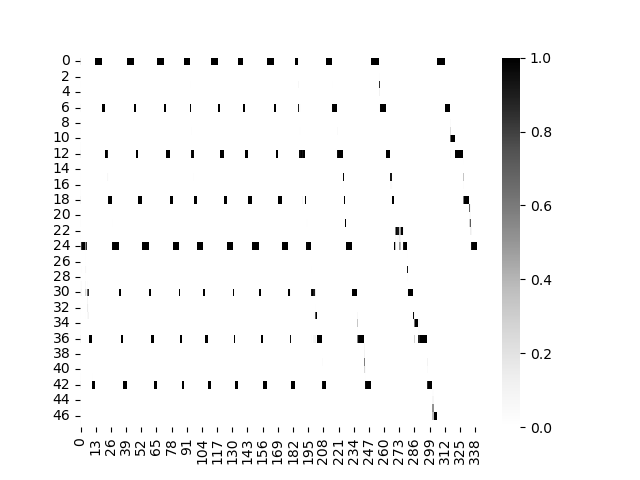}
        \subcaption{Start position}
        \label{fig:start_heat_music}
    \end{minipage}\hfill
    \begin{minipage}[t]{0.32\linewidth}
        \centering
        \includegraphics[width=\linewidth]{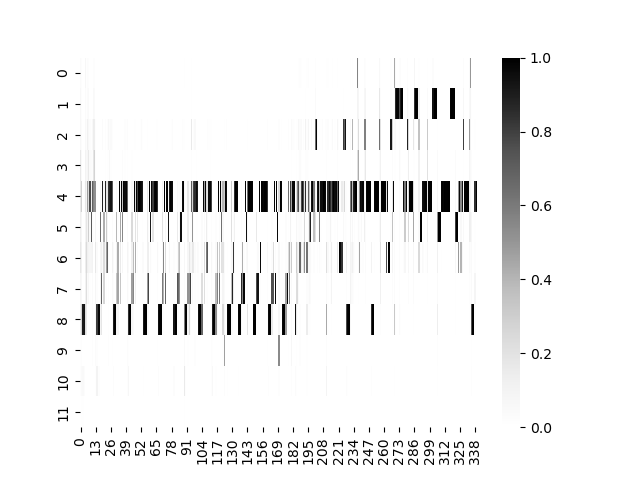}
        \subcaption{Length}
        \label{fig:dur_heat_music}
    \end{minipage}\hfill
    \begin{minipage}[t]{0.32\linewidth}
        \centering
        \includegraphics[width=\linewidth]{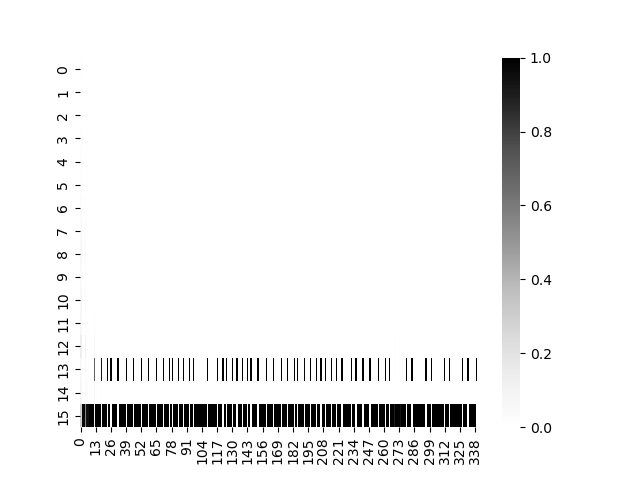}
        \subcaption{Velocity}
        \label{fig:vel_heat_music}
    \end{minipage}
    \caption{Heatmaps during music generation (Musical Attention).}
    \label{fig:heatmap}
\end{figure}

Figure~\ref{fig:heatmap} shows the heatmap for each event when generating the musical composition shown in Figure~\ref{fig:generate}. 
The vertical axis represents the token numbers of each event, while the horizontal axis represents the token index. 
As illustrated in Figure~\ref{fig:inst_heat_music}, it can be observed that the generated music is composed of four instruments: 
guitar (3), bass (4), strings (6), and saxophone (8). 
Figure~\ref{fig:pitch_heat_music} indicates that the notes outside of the C major scale, 
namely D$\flat$ (1), E$\flat$ (3), G$\flat$ (6), A$\flat$ (8), and B$\flat$ (10), have output values close to 0. 
(The numbers 0 through 11 correspond to the 12 pitches from C to B.) 
Figures~\ref{fig:bar_heat_music} and \ref{fig:start_heat_music} show that the music is generated according to the specified number of measures and follows a chronological order. 
Regarding note duration, Figure~\ref{fig:dur_heat_music} shows that the output values vary more than in other events; specifically, eighth notes (4) and half notes (8) exhibit higher values. 
Figure~\ref{fig:vel_heat_music} indicates that the volume is limited to 105 (13) and 115 (15), suggesting that dynamic variation is not well reflected.

\section{Music Theory}
\subsection{Key and Scale} \label{sec:key}

\begin{figure}[h]
    \begin{minipage}[t]{0.49\linewidth}
        \centering
        \includegraphics[width=\linewidth]{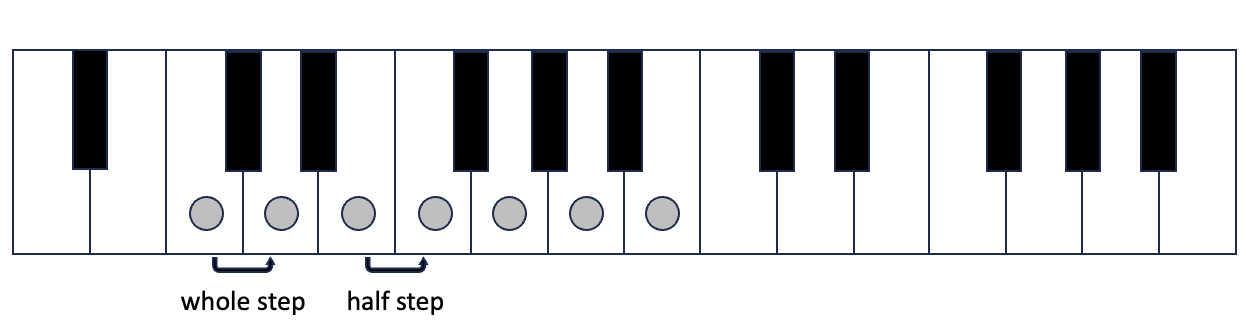}
        \subcaption{C major scale}
    \end{minipage}
    \begin{minipage}[t]{0.49\linewidth}
        \centering
        \includegraphics[width=\linewidth]{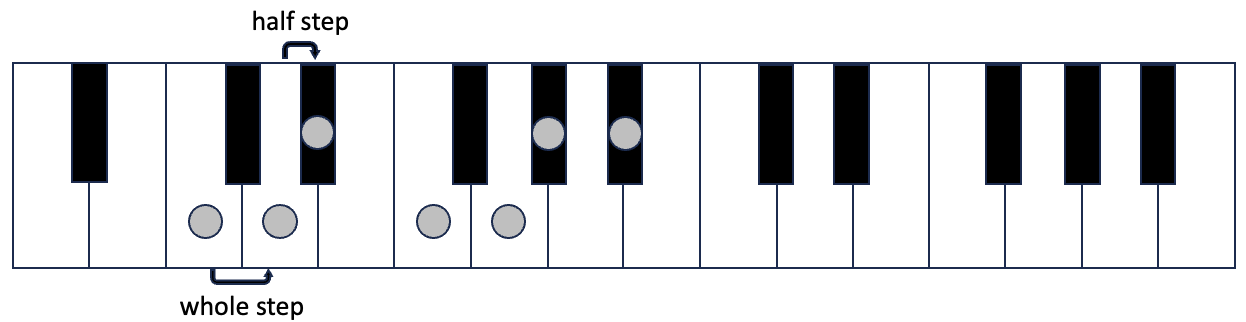}
        \subcaption{C minor scale}
    \end{minipage}
    \caption{
    Examples of the C major scale and the C minor scale.  
    A scale is a set of pitches constructed according to a specific interval pattern based on a reference note (root), which determines the tonal center of a musical piece.  
    The C major scale is generally associated with a bright tonal quality, while the C minor scale is typically perceived as having a darker tonal character.
    }
    \label{fig:key}
\end{figure}

A key refers to the state in which a musical piece is based on a scale with a specific note as the root note.
Scales are classified into major scales and minor scales.
Figure~\ref{fig:key} shows the C major scale and the C minor scale.
The major scale is composed of the following interval pattern from the root note: whole step, whole step, half step, whole step, whole step, whole step, and half step.
The minor scale is composed of: whole step, half step, whole step, whole step, half step, whole step, and whole step.
(A half step refers to a one-note interval, and two half steps make a whole step.)

\subsection{Relative Keys} \label{sec:relative_key}
\begin{figure}[h]
    \begin{minipage}[t]{0.49\linewidth}
        \centering
        \includegraphics[width=\linewidth]{fig/music/C.png}
        \subcaption{C major scale}
    \end{minipage}
    \begin{minipage}[t]{0.49\linewidth}
        \centering
        \includegraphics[width=\linewidth]{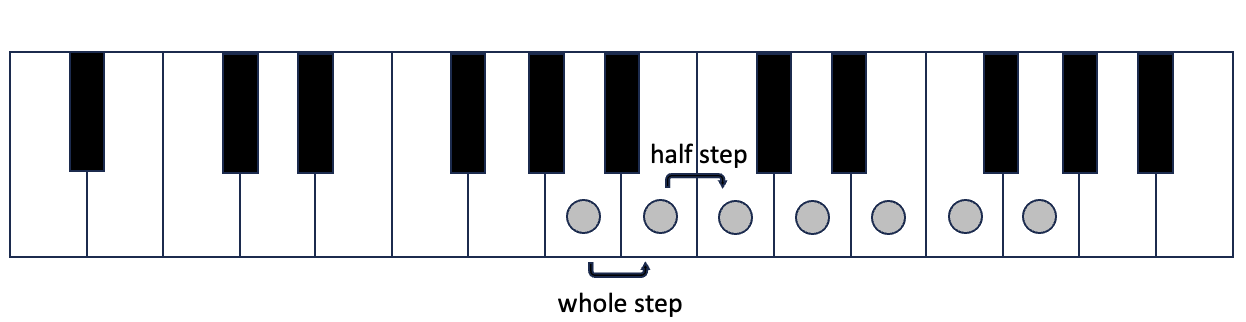}
        \subcaption{A minor scale}
    \end{minipage}
    \caption{Example of relative keys: C major and A minor.
    Relative keys share the same set of scale tones but have different root notes. 
    The change in the tonal center from C to A produces a different musical character while maintaining the same pitch set.}
    \label{fig:relative_key}
\end{figure}

There are cases where the notes comprising the scale are identical in both major and minor keys.
This relationship is called a relative key.
Figure~\ref{fig:relative_key} shows the scales for one such relative key pair: the C major key and the A minor key.

\subsection{Diatonic Chords} \label{sec:diatonic}
\begin{figure}[h]
    \centering
    \includegraphics[width=\linewidth]{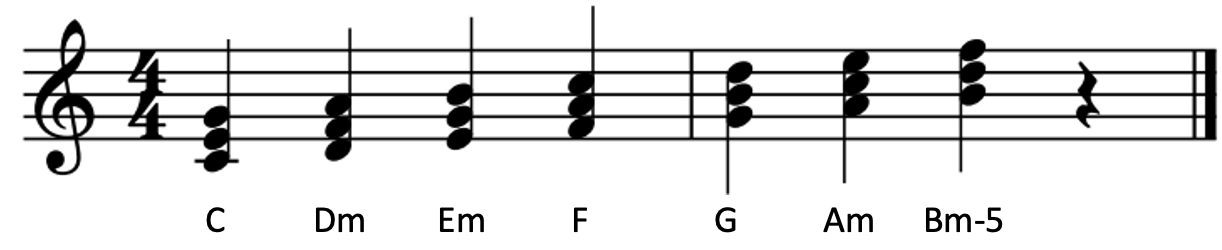}
    \caption{
    Diatonic chords based on the C major scale.
    In the key of C major, there are seven diatonic chords: C, Dm, Em, F, G, Am, and Bm$\flat$5.
    These chords serve as the fundamental units of chord progressions and accompaniment patterns.
    }
    \label{fig:diatonic}
\end{figure}

Diatonic chords are chords constructed from the notes of a specific key.
The notes of the C major scale are C, D, E, F, G, A, and B.
Diatonic triad chords are formed by taking the root, the third, and the fifth degree of each note in the scale.
The diatonic chords in the C major key are C, Dm, Em, F, G, Am, and Bm$\flat$5, as illustrated in Figure~\ref{fig:diatonic}.

\subsection{Arpeggio} \label{sec:arpeggio}
An arpeggio is a method of playing the notes of a chord sequentially, either from the lowest to the highest or vice versa. It is commonly used on instruments such as the guitar or piano to create harmonic and rhythmic movement.


\section{References}
\begin{thebibliography}{99}

    \bibitem{musicvae}Roberts A, Engel J, Raffel C, Simon I, Hawthorne C. MusicVAE: Creating a palette for musical scores with machine learning. arXiv preprint arXiv:1803.05428. 2018.

    \bibitem{midivae}Brunner G, Konrad A, Wang Y, Wattenhofer R. MIDI-VAE: Modeling dynamics and instrumentation of music with applications to style transfer. arXiv preprint arXiv:1809.07600. 2018.

    \bibitem{cyclegan-music}Brunner G, Wang Y, Wattenhofer R, Zhao S. Symbolic music genre transfer with CycleGAN. arXiv preprint arXiv:1809.07575. 2018.

    \bibitem{transformer}Vaswani A, Shazeer N, Parmar N, Uszkoreit J, Jones L, Gomez AN, et al. Attention is all you need. arXiv preprint arXiv:1706.03762. 2017.

    \bibitem{music-bert}Zeng M, Tan X, Wang R, Ju Z, Qin T, Liu TY. MusicBERT: Symbolic music understanding with large-scale pre-training. arXiv preprint arXiv:2106.05630. 2021.

    \bibitem{midi-bert}Chou YH, Chen IC, Chang CJ, Ching J, Yang YH. MidiBERT-Piano: Large-scale pre-training for symbolic music understanding. arXiv preprint arXiv:2107.05223. 2021.

    \bibitem{musicgen}Copet J, Kreuk F, Gat I, Remez T, Kant D, Synnaeve G, et al. Simple and controllable music generation. arXiv preprint arXiv:2306.05284. 2023.

    \bibitem{relative-attention}Shaw P, Uszkoreit J, Vaswani A. Self-attention with relative position representations. arXiv preprint arXiv:1803.02155. 2018.

    \bibitem{music-transformer}Huang CA, Vaswani A, Uszkoreit J, Shazeer N, Simon I, Hawthorne C, et al. Music Transformer: Generating music with long-term structure. arXiv preprint arXiv:1809.04281. 2018.

    \bibitem{musiclm}Agostinelli A, Denk TI, Borsos Z, Engel J, Verzetti M, Caillon A, et al. MusicLM: Generating music from text. arXiv preprint arXiv:2301.11325. 2023.

    \bibitem{mulan}Huang Q, Jansen A, Lee J, Ganti R, Li JY, Ellis DPW. MuLan: A joint embedding of music audio and natural language. arXiv preprint arXiv:2208.12415. 2022.

    \bibitem{midinet}Yang LC, Chou SY, Yang YH. MidiNet: A convolutional generative adversarial network for symbolic-domain music generation. arXiv preprint arXiv:1703.10847. 2017.

    \bibitem{musegan}Dong HW, Hsiao WY, Yang LC, Yang YH. MuseGAN: Multi-track sequential generative adversarial networks for symbolic music generation and accompaniment. arXiv preprint arXiv:1709.06298. 2017.

    \bibitem{gan}Goodfellow IJ, Pouget-Abadie J, Mirza M, Xu B, Warde-Farley D, Ozair S, et al. Generative adversarial nets. arXiv preprint arXiv:1406.2661. 2014.

    \bibitem{bert}Devlin J, Chang MW, Lee K, Toutanova K. BERT: Pre-training of deep bidirectional transformers for language understanding. arXiv preprint arXiv:1810.04805. 2018.

    \bibitem{musenet}Payne C. MuseNet. OpenAI. 2019 Apr 25. Available from: https://openai.com/blog/musenet

    \bibitem{remi}Huang YS, Yang YH. Pop Music Transformer: Beat-based modeling and generation of expressive pop piano compositions. arXiv preprint arXiv:2002.00212. 2020.

    \bibitem{musemorphose}Wu SL, Yang YH. MuseMorphose: Full-song and fine-grained piano music style transfer with one Transformer VAE. arXiv preprint arXiv:2105.04090. 2021.

    \bibitem{sparse-transformer}Child R, Gray S, Radford A, Sutskever I. Generating long sequences with sparse transformers. arXiv preprint arXiv:1904.10509. 2019.

    \bibitem{gpt3}Brown TB, Mann B, Ryder N, Subbiah M, Kaplan J, Dhariwal P, et al. Language models are few-shot learners. arXiv preprint arXiv:2005.14165. 2020.

    \bibitem{instructgpt}Ouyang L, Wu J, Jiang X, Almeida D, Wainwright CL, Mishkin P, et al. Training language models to follow instructions with human feedback. arXiv preprint arXiv:2203.02155. 2022.

    \bibitem{audiolm}Borsos Z, Marinier R, Vincent D, Kharitonov E, Pietquin O, Sharifi M, et al. AudioLM: a language modeling approach to audio generation. arXiv preprint arXiv:2209.03143. 2022.

    \bibitem{soundstream}Zeghidour N, Luebs A, Omran A, Skoglund J, Tagliasacchi M. SoundStream: An end-to-end neural audio codec. arXiv preprint arXiv:2107.03312. 2021.

    \bibitem{wav2bert}Chung YA, Zhang Y, Han W, Chiu CC, Qi J, Pang R, et al. W2v-BERT: Combining contrastive learning and masked language modeling for self-supervised speech pre-training. arXiv preprint arXiv:2108.06209. 2021.

    \bibitem{dataset}Raffel C. Learning-based methods for comparing sequences, with applications to audio-to-MIDI alignment and matching [PhD thesis]. Columbia University; 2016. Available from: https://colinraffel.com/projects/lmd/

    \bibitem{miditoolkit}Yating Music. Miditoolkit: A Python package for working with MIDI files. 2021. Available from: https://github.com/YatingMusic/miditoolkit

\end{thebibliography}
\end{document}